# Pressure-induced spin reorientation transition in layered ferromagnetic insulator Cr$_2$Ge$_2$Te$_6$

Zhisheng Lin[1], Mark Lohmann[1], Zulfikhar A. Ali[1], Chi Tang[1], Junxue Li[1], Wenyu Xing[2], Jiangnan Zhong[2], Shuang Jia[2,3], Wei Han[2,3], Sinisa Coh[4,5], Ward Beyermann[1], and Jing Shi[1]

[1]Department of Physics and Astronomy, University of California, Riverside CA 92521

[2]International Center for Quantum Materials, School of Physics, Peking University, Beijing 100871, P. R. China

[3] Collaborative Innovation Center of Quantum Matter, Beijing 100871, P. R. China

[4]Department of Mechanical Engineering, University of California, Riverside CA 92521

[5]Department of Materials Science and Engineering, University of California, Riverside CA 92521

**ABSTRACT**

Anisotropic magnetoresistance (AMR) of Cr$_2$Ge$_2$Te$_6$ (CGT), a layered ferromagnetic insulator, is investigated under an applied hydrostatic pressure up to 2 GPa. The easy axis direction of the magnetization is inferred from the AMR saturation feature in the presence and absence of the applied pressure. At zero applied pressure, the easy axis is along the *c*-direction or perpendicular to the layer. Upon application of a hydrostatic pressure>1 GPa, the uniaxial anisotropy switches to easy-plane anisotropy which drives the equilibrium magnetization from the *c*-axis to the *ab*-plane at zero magnetic field, which amounts to a giant magnetic anisotropy energy change (>100%). As the temperature is increased across the Curie temperature, the characteristic AMR effect gradually decreases and disappears. Our first-principles calculations confirm the giant magnetic anisotropy energy change with moderate pressure and assign its origin to the increased off-site spin-orbit interaction of Te atoms due to a shorter Cr-Te distance. Such a pressure-induced spin reorientation transition is very rare in three-dimensional ferromagnets, but it may be common to other layered ferromagnets with similar crystal structures to CGT, and therefore offers a unique way to control magnetic anisotropy.

# I. INTRODUCTION

Layered ferromagnets such as $Cr_2Ge_2Te_6$ (CGT), $CrI_3$ and $Fe_3GeTe_2$ have recently received a great deal of attention in the two-dimensional materials community [1-6]. As members of the van der Waals (vdW) material family, they are ideal candidates for both studying fundamental low-dimensional magnetism and fabricating heterostructures with other layered materials for potential spintronics applications. Recently, few layers and even monolayers of these materials have been successfully exfoliated and shown to retain ferromagnetic order at finite temperatures and exhibit fascinating magnetic properties such as the layer number dependence [1, 2]. These studies revealed the importance of both intra-layer and inter-layer interactions. Unique to these layered ferromagnets, the 3d transition metal elements such as Cr and Fe are sandwiched between heavier elements such as Te and I atoms in the outer layers, and the hybridization across the layers imparts a strong spin-orbit interaction in the heavy elements to the 3d transition metal elements, which consequently influences the magnetic anisotropy of the spins in the latter. In spite of the strong shape anisotropy, spins in most layered ferromagnets are aligned perpendicular to the layer, indicative of strong magneto-crystalline anisotropy. In contrast, the exchange interaction is primarily determined by the distance between the 3d transition metal atoms in the same layer. Clearly, understanding the intra- and inter-layer interactions between the spins in 3d transition metal atoms and their interactions with other heavy atoms will shed light on the microscopic origin of their magnetic properties.

Application of pressure offers an effective way of manipulating inter- and intra-layer distances. Even under hydrostatic pressure, the response of the layered ferromagnets is expected to be highly anisotropic since the inter-atomic distances perpendicular to layers are more susceptible to the pressure than those parallel to layers, leading to a highly anisotropic response which is qualitatively different from three-dimensional (3D) ferromagnets. In typical 3D ferromagnets such as Fe, Ni [7], $Fe_3O_4$ [8], the magnetic anisotropy energy (MAE) change under 1 GPa hydrostatic pressure is only about a few percent, too small to cause any spin reorientation transition. In $SrCoO_3$, a Co-based perovskite, a spin reorientation change under 1.1 GPa pressure was suggested by density functional theory calculations [9]. Recent calculations [10] also predict giant pressure induced changes in the MAE of a layered ferromagnet, $Fe_3GeTe_2$. Here, we employ a pressure cell to apply moderate hydrostatic pressure to a CGT crystal up to 2 GPa. Under each fixed pressure, we measure the magneto-transport properties above and below the Curie temperature ($T_c \sim$ 61 K). In its ferromagnetic phase, the magnetoresistance exhibits the characteristic anisotropic behavior with clear saturation above a certain magnetic field, which occurs when the magnetization is fully aligned with the magnetic field. The saturation feature is used to track the magnetization orientation of CGT. We find that the easy axis orientation undergoes a switching from the *c*-axis to *ab*-plane, i.e., spin reorientation transition, upon applying >1 GPa pressure, which represents a giant MAE change compared to conventional 3D ferromagnets. This experimental effect is supported by our first-principles calculations that include Hubbard *U* as well as spin-orbit interaction.

## II.  EXPERIMENTAL SETUP

Single crystalline CGT samples are grown by the flux method [11] and cleaved along the *ab*-plane to platelets of approximately 2 mm×1 mm×0.01 mm. A small amount of indium is first attached to the ends of the CGT sample before gold wires are connected from the sample to the feedthrough wires of the pressure cell for transport measurements. The electrical current flows along the *ab*-plane in the measurements. To ensure good insulation, the gold wires are coated with epoxy (Permatex PermaPoxy) and left in ambient atmosphere for at least 24 hours to fully cure. The sample is then mounted inside a high-pressure cell module (Almax easyLab Pcell 30) that is part of an insert of Quantum Design's Physical Property Measurement System (PPMS). Before any pressure is applied, the sample is immersed in a medium of pentane & iso-pentane with a mix ratio of 1:1. To prevent possible reaction with the medium, the sample is coated with a layer of PMMA. Pressure is applied to the sample through the transmitting medium by exerting a force on a series of tungsten carbide pistons and the pressure is estimated by reading a hydraulic pressure gauge. Due to the sample space limitation, only two wires are connected to the sample and the resistance is measured with this two-terminal geometry. Because of the insulating behavior of CGT itself [11], we expect that two-terminal resistance at low temperatures is dominated by the CGT resistance itself. The magneto-transport measurements are conducted in the PPMS down to 2 K and in a magnetic field up to 10 kOe. The highest pressure we have applied in this work is ~2 GPa. The magnetization measurements are conducted with Quantum Design's Magnetic Property Measurement System.

## III.  RESULTS

Figure 1(a) is an illustration of the crystal structure of CGT containing Te-Ge-Cr-Ge-Te atomic layer units. The distance between the centers of adjacent Ge-Cr-Te layers along [001] is ~0.7 nm, which is larger than other interatomic distances within each unit (e.g., Te-Te: 0.4 nm, Cr-Te: 0.3 nm, Cr-Cr: 0.4 nm, Ge-Te: 0.4 nm or 0.3 nm and Ge-Ge: 0.2 nm)[10]. Similar to other vdW materials, the weak inter-layer interaction allows for exfoliation down to monolayers. In bulk crystals, CGT has a $T_c$ of ~ 61 K and a band gap of ~ 0.2 eV [12]. The magnetic moment of each Cr atom is found to be 2.23 $\mu_B$ from experiments [11]. Figure 1(b) displays the magnetization data measured on a CGT crystal at 5 K when the magnetic field *H* is parallel to the *c*-axis and *ab*-plane. Neither field orientation shows any measurable hysteresis with clear remanent magnetization or coercive field, indicating a soft ferromagnetic behavior. However, the saturation fields are different between these two orientations. When *H* is applied parallel to the *ab*-plane, the saturation field $H_s$ is ~5 kOe, which is larger than that in the direction parallel to the *c*-axis (~ 3 kOe). The anisotropy can be approximately described by a uniaxial term [4].

We carry out magneto-transport measurements in CGT platelet samples across $T_c$. Figure 2(a) plots the magnetoresistance ratio, $\Delta R/R = \frac{R - R_0}{R}$, here *R* and $R_0$ being the two-terminal resistances at finite and zero *H* respectively, measured at 5 K with *H* applied along the *c*-axis or *ab*-plane, i.e., perpendicular or parallel to the platelet, at zero applied pressure. When *H* is applied in the *ab*-plane, it is along the current direction. As already inferred from Figure 1(b), the remanent state at *H*=0 does not have any net magnetization, indicating magnetic moment cancellation due to non-collinear spin configurations such as up and down closure-domain configurations. Hence, when *H* is oriented in the *ab*-plane, the magnetization process is expected to show the typical hard-axis behavior, i.e., the perpendicularly oriented spins rotate towards *H*. The final uniform single-domain state is reached at the saturation field $H_s$ ~ 5 kOe. When the resistance is measured over the same field range, Δ*R*/*R* shows a tendency to saturate, which is observed in Figure 2(b) (curve labelled as "*H*||*ab*"). The saturation behavior in Δ*R*/*R* can be identified by the obvious slope change at ~5 kOe, corresponding well to the magnetization saturation field $H_s$. Δ*R*/*R* saturation suggests the anisotropic magnetoresistance (AMR) origin, i.e., *R* depending on the relative orientation of magnetization direction with respect to current, which is common in conducting ferromagnetic materials [13]. Note that *R* decreases as the magnetization rotates towards the current direction. This AMR response indicates that the resistivity is larger when the magnetization is perpendicular to the current than when they are parallel, i.e., $\rho_\perp > \rho_\parallel$, which has the opposite trend to most ferromagnetic materials [13]. Considering the platelet sample shape, the demagnetizing field is approximately equal to $4\pi M_s$ which is ~ 1.8 kOe. From $H_s = \frac{2K_u}{M_s} - 4\pi M_s$, we estimate the uniaxial anisotropy energy $K_u$ to be ~ $3.65 \times 10^5$ erg/cm$^3$, which agrees with the previously reported value [10].

When *H* is oriented along the *c*-axis, i.e., the easy axis, the magnetizing process is through domain wall motion until the single-domain state is realized. Since *R* is the same between the oppositely oriented spin directions, the single-domain state should have approximately the same resistance as the initial up- and down closure-domain state, providing that the domain-wall magnetoresistance is neglected, which is the case in most ferromagnets. Therefore, *R* should not change between the initial and final magnetization configurations from the AMR mechanism, and the observed smooth Δ*R*/*R* curve is merely the magnetic field $H$-dependent background, as shown in Figure 2(b) (curve labelled as "$H||c$"). This negative magnetoresistance background may be simply caused by suppressed spin-flip scattering at high magnetic fields. Clearly, the $H||ab$ curve also contains a similar $H$-dependent magnetoresistance background which can be seen in the high field region ($H > 5$ kOe). In order to separate both effects, we perform polar angular dependence measurements by rotating the sample. As the CGT sample rotates in a constant magnetic field (10 kOe), i.e., $\theta_{xz}$ from 0 to 90° as illustrated in Figure 2(a), the magnetization remains aligned with the fixed field direction but rotates relatively to the current. Thus, this experiment only tracks the magnetization orientation dependence with respect to the current direction, i.e., only AMR. The angular dependence can be well fitted with $\sin^2 \theta_{xz}$ [14], as shown in Figure 2(c). Note that $\Delta R/R$ is larger at $\theta_{xz} = 0$ than $\theta_{xz} = 90°$, which is consistent with $\rho_\parallel < \rho_\perp$. The AMR signal rapidly decreases as the temperature approaches $T_c$, but it remains finite even above $T_c$ due to aligned paramagnetic spins. The slightly smaller $\Delta R/R$ at 5 K compared to 15 K could be due to the contact resistance which adds in serial to the two-terminal resistance.

The pressured-induced AMR changes suggest more profound electronic property modifications due to the highly anisotropic atomic arrangements in layered materials. The electronic property changes, ranging from quantitative bandgap reduction to semiconductor-semimetal transition, occur in other layered materials such as black phosphorus, graphene and transition metal dichalcogenides [15-18]. To study the electronic properties of CGT, we measure the resistance vs. temperature under 0 and 2 GPa. The two-terminal $R$ increases as the temperature is decreased, which is consistent with the previously reported insulating behavior [11]. Upon application of hydrostatic pressure, $R$ decreases at room temperature. The difference in $R$ between 0 and 2 GPa enlarges dramatically as the temperature is decreased. Figure 3 shows the temperature dependence of $R$ under 0 GPa and 2 GPa from 100 to 280 K. Although $R$ starts from approximately the same value at 280 K, at 100 K under 2 GPa, it is reduced to ~ 1/20 of the 0 GPa value. Over this temperature range, $R$ in both cases can be fitted by the activated behavior with an energy gap $E_g$ described by the Arrhenius equation, $R = R_0 \exp(\frac{E_g}{2k_B T})$, yielding $E_g$ of 0.19 eV at 0 GPa and 0.12 eV, decreased by 36.8% at 2 GPa. The 0 GPa bandgap is consistent with the previously reported value [11, 19]. Below 100 K, the increasing trend of $R$ continues, but the rate of the increase becomes smaller and $R$ deviates from the simple Arrhenius equation suggesting the variable range hopping. Hence we only model the pressure effect on the high-temperature $R$ behavior by a decreased bandgap.

To study the effect on magnetoresistance, we apply a series of pressures to perform the $\Delta R/R$ measurements with $H$ applied along the $c$-axis under pressure below 80 K. Let us first examine the comparison between 0 GPa and 2 GPa at 15 K is shown in Figure 4(a). There are two marked differences. First, although there is no saturation behavior in the 0 GPa magnetoresistance as discussed earlier, under 2 GPa, a clear saturation feature emerges as $H$ approaches ~6 kOe. It indicates that under 2 GPa the initial magnetization orientation is not parallel or anti-parallel to the applied field direction as in the 0 GPa case, i.e., no longer along the *c*-axis. The new initial magnetization state indicates suppressed uniaxial anisotropy with a reduced or negative $K_u$ term. In fact, it requires a ~ 6 kOe *c*-axis field to fully rotate the magnetization as shown by the saturation in $\Delta R/R$. This field is larger than the demagnetizing field, 1.8 kOe, which means a pressure-induced negative $K_u$ term or easy-plane anisotropy, or a pressured-induced spin reorientation transition. By comparing the saturation fields between 0 and 2 GPa with the consideration of the same demagnetizing field, we estimate that the pressure-induced easy-plane MAE magnitude is about 1.6 times as large as the *c*-axis uniaxial MAE at 0 GPa. Second, since $H$ is oriented out-of-plane, at high fields, the magnetization is perpendicular to the current, and the resistivity due to AMR approaches $\rho_\perp$. The apparent sign reversal from negative to positive magnetoresistance upon 2 GPa pressure does not mean any change in resistivity anisotropy. On contrary, because of the different initial magnetization orientations, the magnetoresistance sign reversal confirms that the same $\rho_\parallel < \rho_\perp$ relation holds under pressure. From these observations, we conclude that easy-plane anisotropy emerges under 2 GPa which is sufficiently strong to destabilize the *c*-axis spin orientation at $H$=0. As illustrated in Figure 4(a), this transition occurs between 1.0 to 1.5 GPa.

As shown in Figures 4(b), in the absence of pressure, $\Delta R/R$ under the *c*-axis field is negative at all temperatures. It is consistent with the perpendicular magnetic anisotropy at 0 GPa. The origin of the negative $\Delta R/R$ is no different from the $H||c$ curve in Figure 2(b). Under 2.0 GPa, however, spin reorientation transition occurs and the initial magnetization is in the *ab*-plane below the ferromagnetic transition temperature. Such a transition results in positive $\Delta R/R$ in the small field range followed by saturation of $\Delta R/R$ as shown in Figure 4(c) at two representative temperatures. The positive $\Delta R/R$ decreases as the temperature increases. At 80 K, $\Delta R/R$ becomes negative and the saturation feature seems to disappear, but it is clearly different from the 0 GPa 80 K curve in Figure 4(b). At 1.5 GPa, the evolution of the magnetoresistance effect is similar. It is worth noting that the 2 GPa 80 K magnetoresistance more closely resembles the 0 GPa 50 K data than the 0 GPa 80 K data, which suggests a possible $T_c$ enhancement under pressure.

Our first-principles calculations, using Quantum-ESPRESSO package, [20, 21] on bulk CGT at both zero and finite pressure reveal the origin of the pressure-induced change in MAE. We find that upon inclusion of the Hubbard U interaction of 1.0 eV the MAE, $\Delta E = E_c - E_{ab}$, of CGT at 0 GPa equals -77 μeV per Cr atom, i.e. magnetization orientation prefers the *c*-axis over the *ab*-plane. This anisotropy, as well as choice of Hubbard *U*, is in good agreement with earlier calculations from Ref. [1]. Next, we fully relax the structure under hydrostatic pressure and recompute the MAE. We find that at 2.5 GPa it goes through zero and has an opposite sign for larger pressures. Therefore, above 2.5 GPa, magnetization prefers to point in the *ab*-plane, consistent with our experimental finding. The exact value of the calculated transition pressure depends on the value of the Hubbard *U*.

The MAE calculated above originates purely from the relativistic effects such as spin-orbit interaction. One can easily show that without the relativistic effects, $\Delta E = 0$. Within the commonly used pseudopotential approach [22] to the density functional theory we treat the relativistic effects by solving the Dirac equation for an isolated Cr, Ge, and Te atom and use the corresponding relativistic potential in the bulk calculations for CGT. This approach also enables us to selectively turn on or off the relativistic effects on individual atoms in the calculation.

Since the magnetization of CGT is dominantly from Cr spins, one might think that relativistic effects of Cr atoms also dominate $\Delta E$. However, we find that not to be the case. If we repeat the earlier calculations but with relativistic effects only on Cr atoms, we find that the $\Delta E$ is negligibly small (only -3 µeV). Clearly, an important contribution to MAE must come from the relativistic effects on Ge or Te atoms. To differentiate the influence of each atom, we first recompute $\Delta E$ when relativistic effects are included only on Cr and Te atoms. In this case we get $\Delta E$ =-122 µeV. On the other hand, if we include relativistic effects on Cr and Ge we get $\Delta E$ = 50 µeV. Therefore, we conclude that at zero pressure the relativistic effects originating from Cr and Te atoms tend to prefer Cr-spins aligned in the *c*-axis, while Cr & Ge atoms prefer them in the *ab*-plane.

If we now redo these calculations with applied pressure, we find that the Cr and Te contribution is significantly changed from -122 µeV at 0 GPa to -31 µeV at 2.5 GPa. On the other hand, the Cr and Ge contributions are only slightly modified from 50 µeV at 0 GPa to 62 µeV at 2.5 GPa. Therefore, the pressure dependent relativistic effects originating from Te atoms have a decisive role in changing the sign of the MAE in CGT.

The importance of Te atoms on MAE is further corroborated by analyzing the crystal structure of CGT under applied hydrostatic pressure. If we compare the calculated crystal structure of CGT at 0 GPa and 2.5 GPa, we find that the thickness of each CGT unit is reduced from 3.308 Å at 0 GPa to 3.258 Å at 2.5 GPa, or by 1.5%. Here we define thickness of each unit as the vertical distance between Te atoms in the same CGT unit. If we now repeat the MAE calculation for the structure where instead of applying hydrostatic pressure we only change the vertical Te coordinate so that the CGT unit thickness is kept at 3.258 Å (same as at 2.5 GPa), we find that the MAE is nearly the same as if we used the fully relaxed structure at 2.5 GPa. Therefore, we conclude that the dominant effect of pressure on MAE originates from the vertical displacement of Te atoms. We also confirmed that keeping the thickness of each unit the same, but changing the distance between the CGT layers, does not affect MAE. This off-site spin-orbit interaction of Cr and Te atoms and modification of Cr-Te distance with pressure differentiate the microscopic origin of the giant pressure-induced MAE in CGT from other 3D ferromagnets such as Fe, Ni [7], $Fe_3O_4$ [8], etc.

## IV. DISCUSSION

With the application of a moderate hydrostatic pressure, we have observed interesting effects in a bulk CGT single crystal on both its electronic and magnetic properties. The former can be modeled by a reduced band gap which produces a factor of ~1/20 reduction in resistivity at 100 K with 2 GPa pressure. The latter effect is represented by a more than 100% change in MAE which drives the spin reorientation transition from the *c*-axis to *ab*-plane. From the first-principles calculations, we have identified that the vertical Cr-Te distance reduction within the CGT unit is primarily responsible for the giant MAE change. The threshold of the spin reorientation transition is a < 1.5% reduction in the single CGT unit thickness, which represents a giant magnetostriction effect. Such an extraordinary pressure-induced MAE change distinguishes CGT from most conventional 3D ferromagnets. On the atomic scale, we have found that the off-site spin-orbit interaction between Cr-Te and the highly anisotropic atomic displacements are the fundamental cause. These properties may be common among other layered ferromagnets. Our study indicates a possibility of controlling the preferred spin orientation even down to monolayer thick materials by applying moderate pressure or engineering strain.

## ACKNOWLEDGEMENTS

This work was supported by a DOE BES Award No. DE-FG02-07ER46351. We thank Yadong Xu and Zhong Shi for their technical assistance.

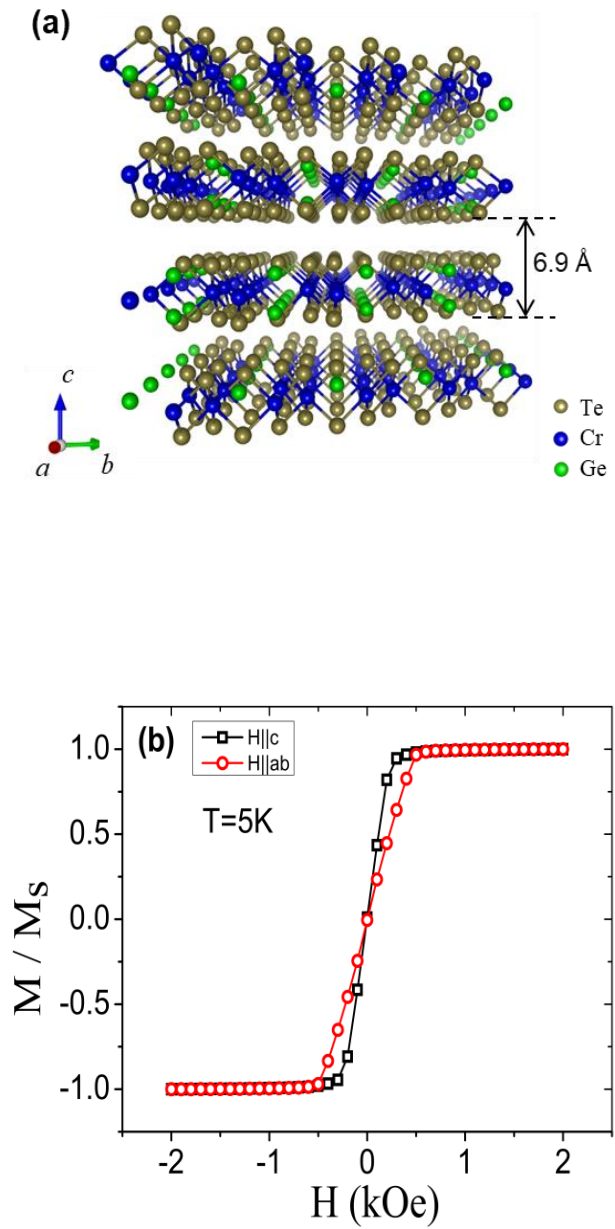

**Figure 1.** (a) Crystal structure of CGT viewed from the *a*-axis direction. (b) Magnetization of a CGT crystal measured at 5 K with the magnetic field parallel to the *c*-axis and *ab*-plane. Both hysteresis loops go through zero within the experimental uncertainty of the superconducting magnet (~ 10 Oe).

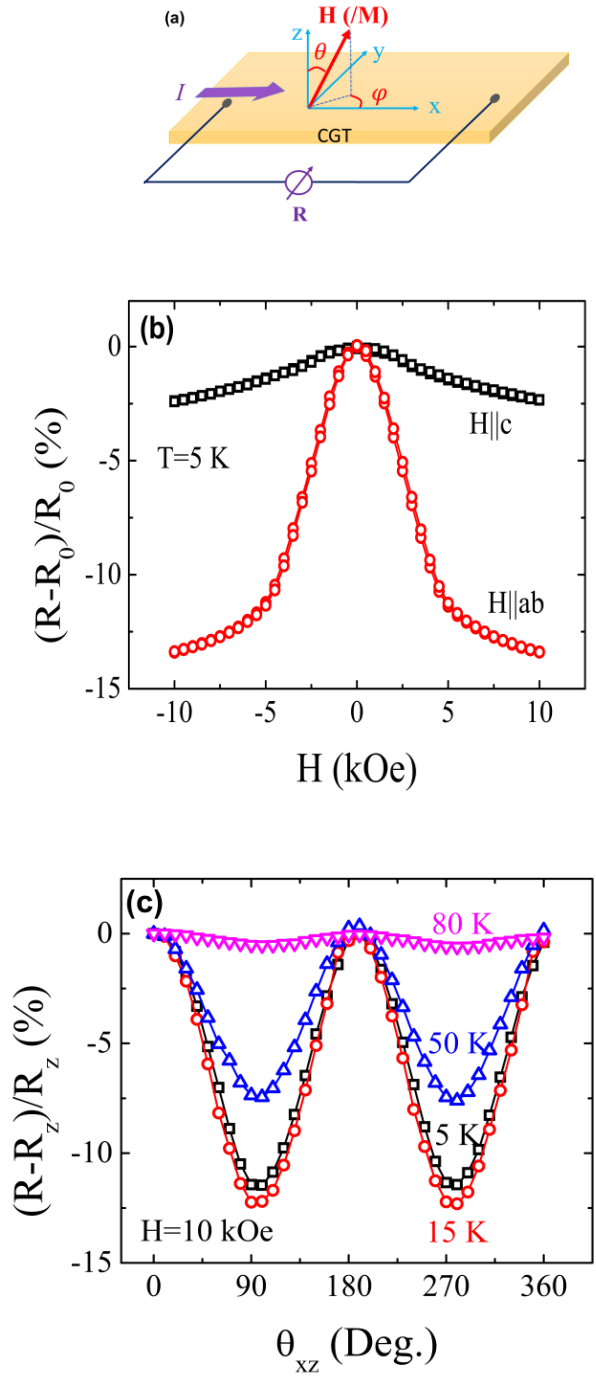

**Figure 2.** (a) Transport measurement geometry. (b) Comparison of magnetoresistance measured at 5 K at 0 GPa with magnetic field along the *c*-axis and *ab*-plane. (c) Angle-dependent magnetoresistance at different temperatures (5, 15, 50 and 80 K) at 0 GPa. Magnetic field orientation in the *xz*-plane starts from out-of-plane ($\theta_{xz}=0°$), through in-plane ($\theta_{xz}=90°$), and completes 360°.

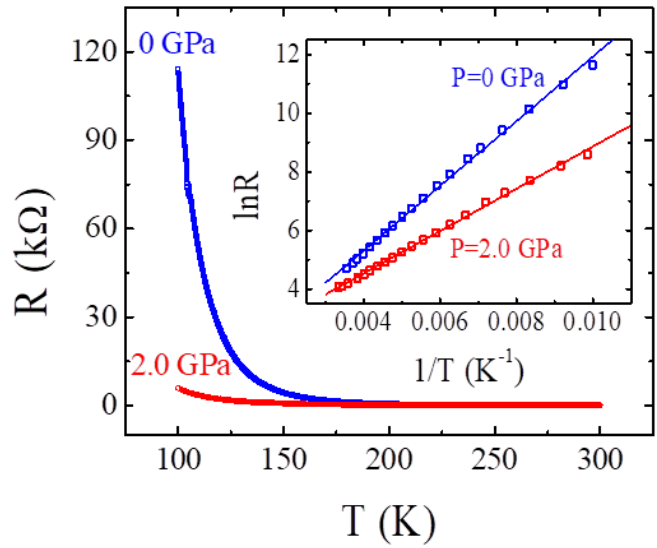

**Figure 3.** Temperature dependence of two-terminal resistance under 0 GPa and 2 GPa from room temperature to 100 K. Inset shows the same data plotted as lnR vs. 1/T.

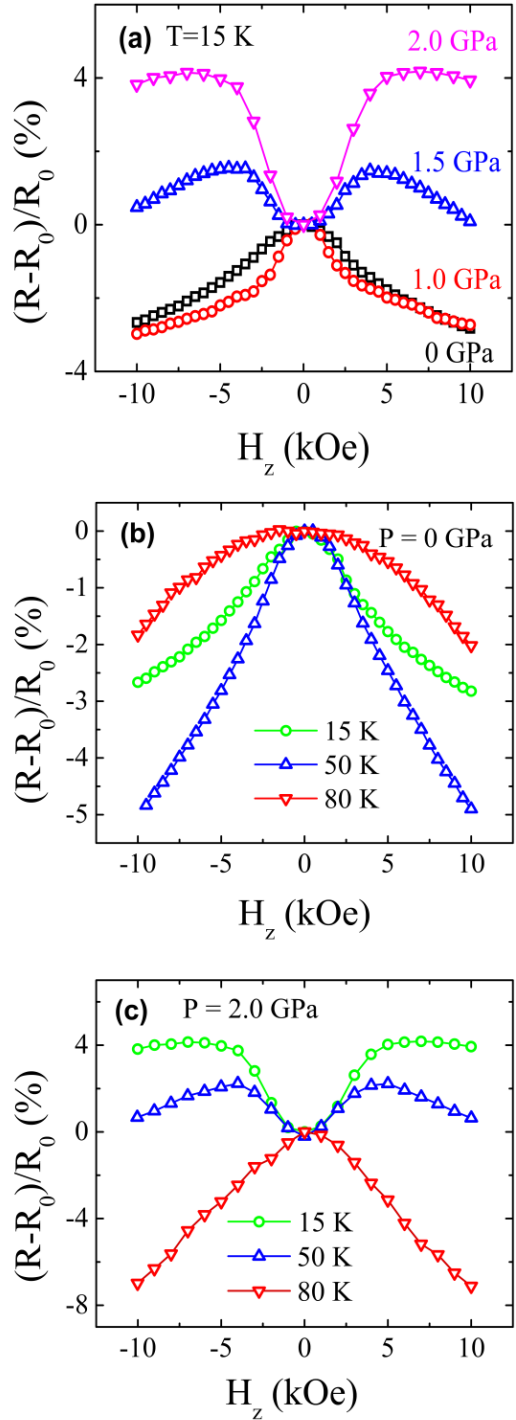

**Figure 4.** (a) Comparison of 15 K magnetoresistance data among 0 GPa, 1.0 GPa, 1.5 GPa, and 2 GPa. Magnetoresistance ratio under 0 GPa (b) and 2 GPa (c) was measured at 15 K, 50 K, and 80 K. Magnetic field is applied along the *c*-axis in all cases.

**References**

[1] C. Gong, L. Li, Z. Li, H. Ji, A. Stern, Y. Xia, T. Cao, W. Bao, C. Wang, Y. Wang, Z. Q. Qiu, R. J. Cava, S. G. Louie, J. Xia, and X. Zhang, Discovery of intrinsic ferromagnetism in two-dimensional van der Waals crystals, Nature **546**, 265-9 (2017).

[2] B. Huang, G. Clark, E. N. Moratalla, D. R. Klein, R. Cheng, K. L. Seyler, D. Zhong, E. Schmidgall, M. A. McGuire, D. H. Cobden, W. Yao, D. Xiao, P. J. Herrero, and X. Xu, Layer-dependent Ferromagnetism in a van der Waals Crystal down to the Monolayer Limit, Nature **546**, 270-3 (2017).

[3] N. Samarth, Condensed-matter physics: Magnetism in flatland, Nature **546**, 216-8 (2017).

[4] W. Xing, Y. Chen, P. M. Odenthal, X. Zhang, W. Yuan, T. Su, Q. Song, T. Wang, J. Zhong, S. Jia, X. C. Xie, Y. Li, and W. Han, Electric Field Effect in Multilayer $Cr_2Ge_2Te_6$: a Ferromagnetic Two-Dimensional Material, 2D Mater. **4**, 024009 (2017).

[5] Y. Wang, C. Xian, J. Wang, B. Liu, L. Ling, L. Zhang, L. Cao, Z. Qu, and Y. Xiong, Anisotropic anomalous Hall effect in triangular itinerant ferromagnet $Fe_3GeTe_2$, Phys. Rev. B **96**, 134428 (2017).

[6] B. Chen, J. Yang, H. Wang, M. Imai, H. Ohta, C. Michioka, K. Yoshimura, and M. Fang, Magnetic properties of layered itinerant electron ferromagnet $Fe_3GeTe_2$, Journal of the Physical Society of Japan **82**, 124711 (2013).

[7] A. Sawaoka and N. Kawai, The effect of hydrostatic pressure on the magnetic anisotropy of ferrous and ferric ions in ferrites with spinel structure, Journal of the Physical Society of Japan **25**, 133-40 (1968).

[8] A. Sawaoka and N. Kawai, Change of the magnetic anisotropy constant K1 of magnetite ($Fe_3O_4$) under hydrostatic pressure, Phys. Lett. A **24**, 503-4 (1967).

[9] J. Yang, C. Terakura, M. Medarde, J. White, D. Sheptyakov, X. Yan, N. Li, W. Yang, H. Xia, J. Dai, Y. Yin, Y. Jiao, J. Cheng, Y. Bu, Q. Zhang, X. Li, C. Jin, Y. Taguchi, Y. Tokura, and Y. Long, Pressure-induced spin reorientation and spin state transition in $SrCoO_3$, Phys. Rev. B **92**, 195147 (2015).

[10] H. L. Zhuang, P. R. C. Kent, and R. G. Hennig, Strong anisotropy and magnetostriction in the two-dimensional Stoner ferromagnet $Fe_3GeTe_2$, Phys. Rev. B **93**, 134407 (2016).

[11] X. Zhang, Y. Zhao, Q. Song, S. Jia, J. Shi, and W. Han, Magnetic anisotropy of the single-crystalline ferromagnetic insulator $Cr_2Ge_2Te_6$, Japanese Journal of Applied Physics **55**, 3 (2016).

[12] V. Carteaux, D. Brunet, G. Ouvrard, and G. Andre, Crystallographic, magnetic and electronic structures of a new layered ferromagnetic compound $Cr_2Ge_2Te_6$, Journal of Physics: Condensed Matter **7**, 69 (1995).

[13] T. McGuire and R. L. Potter, Anisotropic magnetoresistance in ferromagnetic 3d alloys, IEEE Transactions on Magnetics **11**, 1018-38 (1975).

[14] T. Lin, C. Tang, H. M. Alyahayaei, and J. Shi, Experimental investigation of the nature of the magnetoresistance effects in Pd-YIG hybrid structures, Phys. Rev Lett. **113**, 037203 (2014).

[15] Z. J. Xiang, G. J. Ye, C. Shang, B. Lei, N. Z. Wang, K. S. Yang, D. Y. Liu, F. B. Meng, X. G. Luo, L. J. Zou, and Z. Sun, Pressure-induced electronic transition in black phosphorus, Phys. Rev. Lett. **115**, 186403 (2015).

[16] J. E. Proctor, E. Gregoryanz, K. S. Novoselov, M. Lotya, J. N. Coleman, and M. P. Halsall, High-pressure Raman spectroscopy of graphene, Phys. Rev. B **80**, 073408 (2009).

[17] A. P. Nayak, S. Bhattacharyya, J. Zhu, J. Liu, X. Wu, T. Pandey, C. Jin, A. K. Singh, D. Akinwande, and J. F. Lin, Pressure-induced semiconducting to metallic transition in multilayered molybdenum disulphide, Nature communications **5**, 3731 (2014).

[18] Z. H. Chi, X. M. Zhao, H. Zhang, A. F. Goncharov, S. S. Lobanov, T. Kagayama, M. Sakata, and X. J. Chen, Pressure-induced metallization of molybdenum disulfide, Phys. Rev. Lett. **113**, 036802 (2014).

[19] H. Ji, R. A. Stokes, L. D. Alegria, E. C. Blomberg, M. A. Tanatar, A. Reijnders, L. M. Schoop, T. Liang, R. Prozorov, K. S. Burch, and N. P. Ong, A ferromagnetic insulating substrate for the epitaxial growth of topological insulators, Journal of Applied Physics **114**, 114907 (2013).

[20] K. F. Garrity, J. W. Bennett, K. M. Rabe, and D. Vanderbilt, Pseudopotentials for high-throughput DFT calculations, Computational Materials Science **81**, 446-52 (2014).

[21] P. Giannozzi, S. Baroni, N. Bonini, M. Calandra, R. Car, G. Cavazzoni, D. Ceresoli, G. L. Chiarotti, M. Cococcioni, and I. Dabo, QUANTUM ESPRESSO: a modular and open-source software project for quantum simulations of materials, Journal of physics: Condensed matter **21**, 395502 (2009).

[22] L. Kleinman, Relativistic norm-conserving pseudopotential, Phys. Rev. B **21,** 2630 (1980).